\documentclass[twocolumn,showpacs,showkeys]{revtex4}
\usepackage{graphicx}
\usepackage{bm}
\usepackage{color}
\usepackage{amsmath}
\usepackage{natbib}

\begin{document}


\textbf{Comment on "Spin Contribution to the Ponderomotive Force in a Plasma"}

\textbf{Pavel A. Andreev}

\emph{andreevpa@physics.msu.ru}

\textit{Lomonosov Moscow State University, Moscow, Russia.}

Several years ago, G. Brodin, A. P. Misra, and M. Marklund considered "Spin Contribution to the Ponderomotive Force in a Plasma" \cite{Brodin PRL 10 SPF}. They applied a two fluid model of electron gas where spin-up and spin-down electrons are considered as two different species. For each species of electrons they used the continuity equations, the Euler equations and the spin evolution equations coinciding with the equations obtained for single fluid description of electrons. This approach, however, appears to be incorrect.

Single fluid quantum model of electrons results from the Pauli equation \cite{Takabayasi PTP 55 b}, or, more precisely, from the many-particle Pauli equation \cite{MaksimovTMP 2001}, \cite{Andreev RPJ 07}.

The single particle Pauli equation presents the evolution of the two-component spinor wave function $\psi$ describing the probability of the electron to have a spin directed "up" or "down" (parallel or anti-parallel to the external magnetic field). A similar, but rather more complicate, situation can be observed in many-particle systems of spin-1/2 particles. The fluidisation of the many-particle model \cite{MaksimovTMP 2001} leads to a model which, under the self-consistent field approximation and with no explicit account of thermal evolution, such as time evolution of the energy density or pressure, reveals in a set of equations similar to the single particle case \cite{Takabayasi PTP 55 b}. Thus, if we want to capture the main properties of the two-fluid (separated spin-up and spin-down) model of electrons we can use the single particle Pauli equation.

If we want to trace the evolution of spin-up and spin-down electrons separately we should consider $ \psi_{u} $ and $\psi_{d}$ separately as well. Each of the one component wave functions defines the concentration of electrons $n_u=\psi_{u}^* \psi_{u} $ and $n_d=\psi_{d}^* \psi_{d} $, whereas the full concentration $n= n_u + n_d=\psi^+ \psi $. Starting with these definitions of concentrations for spin-up and spin-down electrons we can perform a two-fluid fluidization of the Pauli equation.

The numbers of spin-up and spin-down electrons are conserved in the absence of the spin-spin interaction. The spin-spin interaction leads to the nontrivial evolution of the spin densities $S_x$ and $S_y$. In this comment we do not consider the effects related to the nonconservation of numbers of spin-up and spin-down electrons.

While the z-projection of spin density $S_{z}$ of electrons is not an independent variable in this model, $S_{z}$ appears as the difference between the concentrations of electrons with different projections of spin $S_{z}=n_{u}-n_{d}$ due to its definition $S_{z}=\psi^{+}\sigma_{z}\psi$.

The spin-up and spin-down directions are related to a preferable direction in space. If we have a strong uniform external magnetic field, its direction can be taken as a preferable direction. This field reveals in z-projection of the magnetic field $B_{z}$ in the Pauli equation. However, the propagation of an electromagnetic wave parallel to z axis creates $B_{x}$ and $B_{y}$ as well.

Considering the time evolution of the probability densities $n_{u}$ and $n_{d}$ we derive the continuity equations
$\partial_{t}n_{s}+\nabla\cdot(n_{s}\textbf{v}_{s})=0 $.
They show the conservation of the particle number. The Euler equation arises as
$$(\partial_{t}+\textbf{v}_{s}\cdot\nabla)\textbf{v}_{s}+\frac{\nabla p_{s}}{mn_{s}}-\frac{\hbar^{2}}{2m^{2}}\nabla\biggl(\frac{\triangle \sqrt{n_{s}}}{\sqrt{n_{s}}}\biggr)=\frac{q_{e}}{m}\biggl(\textbf{E}+\frac{1}{c}[\textbf{v}_{s},\textbf{B}]\biggr)$$
\begin{equation}\label{SUSDC Euler eq electrons spin UP} +(-1)^{i_{s}}\frac{\mu}{m}\nabla B_{z} +\frac{\mu}{2mn_{s}}(S_{x}\nabla B_{x}+S_{y}\nabla B_{y}), \end{equation}
where $q_{e}=-e$, $\mu=-\mu_{B}$, with $\mu_{B}=e\hbar/(2mc)$ being the Bohr magneton, and $i_{u}=0$, $i_{d}=1$.

We have used the notations $S_{x}=\psi^{*}\sigma_{x}\psi$ and $S_{y}=\psi^{*}\sigma_{y}\psi$ in the equation (\ref{SUSDC Euler eq electrons spin UP}). $S_{x}$ and $S_{y}$ appear as mixed combinations of $\psi_{u}$ and $\psi_{d}$. $S_{x}$ and $S_{y}$ describe the simultaneous evolution of both species and do not wear subindexes $u$ and $d$.
The spin evolution is described by $\partial_{t}S^{\alpha}=\frac{2\mu}{\hbar}\varepsilon^{\alpha\beta\gamma}S^{\beta}B^{\gamma}$,
where $\alpha$ stands for $x,y$, whereas $\beta,\gamma=x,y,z$, and $S_{z}=n_{u}-n_{d}$.

We found that the ponderomotive forces acting on spin-up and spin-down electrons
\begin{equation}\label{SUSDC PonForce} \textbf{F}_{s}=\pm\frac{\mu^{2}}{2\hbar}\frac{n_{u}-n_{d}}{n_{s}}\frac{\partial_{z}\mid B_{\pm}\mid^{2}}{\omega\mp\Omega}\end{equation}
have the same sign for the spin-up and the spin-down electrons. This force depends on the polarisation of the light and the rate of the spin polarisation of the medium. The module of the force is found to be four times smaller than the result of the Ref. \cite{Brodin PRL 10 SPF}.

We conclude that Brodin et al. \cite{Brodin PRL 10 SPF} started with the incorrect model of the separate description of the spin-up and the spin-down electrons. As a result, they obtained the incorrect generalisation for the ponderomotive force. The correct model and the correct expression for the ponderomotive force of the zeroth order are presented in this comment.


\end{document}